\documentclass[MS]{macro/neu_msthesis}

\title{Designing Mixed-Initiative Video Games}

\author{Daijin Yang}

\dept{College of Art, Media, and Design}

\degreename{Game Science and Design}

\field{Game Science and Design}

\submitdate{April 2023}

\numberofmembers{3}

\principaladviser{Dr. Elina Tochilnikova}
\firstreader{First Reader Name}
\secondreader{Second Reader Name}

\thirdreader{Third Reader Name}
\fourthreader{Fourth Reader Name}
\fifthreader{Fifth Reader Name}

\chairman{Dr. Shelia Hemami}

\dean{Dr. Sara Wadia-Fascetti}

\usepackage{amsfonts}			
\usepackage{times}		

\newcommand{\ifno}[1]{}

\usepackage{multirow}
\makeindex
\usepackage{makeidx}
\usepackage{acronym}

\usepackage{url}
\ifnum\pdfoutput>0
\usepackage[pdftex]{graphicx}
\else
\usepackage{graphicx}
\fi

\ifnum\pdfoutput>0
\usepackage[pdftex,
bookmarks=true,
bookmarksnumbered=true,
hypertexnames=false,
breaklinks=true           
]{hyperref}
\else
\usepackage[hypertex,
bookmarks=true,
bookmarksnumbered=true,
hypertexnames=false,
breaklinks=true           
]{hyperref}
\fi

\hypersetup{				
pdfauthor = {\authorRef},
pdftitle = {\titleRef},
pdfsubject = {\expandafter{\degreeRef} thesis submitted to Northeastern University},
pdfkeywords = {add keywords here}
pdfcreator = {LaTeX with hyperref package},
pdfproducer = {dvips + ps2pdf}}

\begin{document}

\pdfbookmark[1]{Cover}{cover}

\titlepage

\begin{frontmatter}

\begin{dedication}
To my family.
\end{dedication}

\pdfbookmark[1]{Table of Contents}{contents}
\tableofcontents
\listoffigures
\newpage\ssp
\listoftables



\begin{acknowledgements}

I would like to first thank my great, beautiful, and strong mother. Since being diagnosed with multiple myeloma nine years ago, she has endured unimaginable suffering, but she has never given up on her life and has even achieved more in her work than healthy people. After contracting COVID-19, her condition deteriorated rapidly, and as I write this paper, she is fighting bravely against cancer in the hospital. She has always inspired and supported me to move forward. She is a great mother and I will love her forever.

I am grateful to my father, my mother's sister, and other family members for taking care of my mother and allowing me to focus on my studies.

I would thank Professor Elina Tochilnikova, Professor Giovanni Maria Troiano, Professor Bob De Schutter, Professor Casper Harteveld, Professor Leanne Chukoskie, and all other professors in the field of Game Science and Design at Northeastern University for their invaluable guidance and unwavering patience in supporting my work. I would also express my sincere gratitude to Professor Max Kreminski at Santa Clara University for providing crucial feedback and suggestions on my thesis.

I would like to extend my appreciation to all of my colleagues who generously provided valuable suggestions and constructive feedback on my work. Additionally, I am grateful to my friends Binyao Jian and Xinyan Deng, who stood by me during the most challenging times. Their unwavering support and companionship have been invaluable to me.

\end{acknowledgements}


\begin{abstract}
The development of Artificial Intelligence (AI) enables humans to co-create content with machines. The unexpectedness of AI-generated content can bring inspiration and entertainment to users. However, the co-creation interactions are always designed for content creators and have poor accessibility. To explore gamification of mixed-initiative co-creation and make human-AI interactions accessible and fun for players, I prototyped \textit{Snake Story}, a mixed-initiative game where players can select AI-generated texts to write a story of a snake by playing a “\textit{Snake}” like game. A controlled experiment was conducted to investigate the dynamics of player-AI interactions with and without the game component in the designed interface. As a result of a study with 11 players (n=11), I found that
players utilized different strategies when playing with the two versions, 
game mechanics significantly affected the output stories, players' creative process, as well as role perceptions, 
and players with different backgrounds showed different preferences for the two versions. 
Based on these results, I further discussed considerations for mixed-initiative game design.
This work aims to inspire the design of engaging co-creation experiences.


\end{abstract}

\end{frontmatter}


\pagestyle{headings}


\chapter{Introduction}
\label{chap:intro}

Recent machine learning (ML) techniques have boosted human creation, enabling humans to co-work with artificial intelligence (AI) to compose music \cite{music-cocreation,music-cocreation2,musiccocreation3}, draw illustrations \cite{drawingcocreation,drawingcocreation2,drawingcocreation3}, write stories \cite{cowriteedit1,cowritingfillgap,co-writing-fromtooltocompanion,cowriting-wordcraft,cowritingideageneration1}, reply emails \cite{cowritingemail}, create characters \cite{cowritecreatecharacter}, and develop games \cite{gamedesigncocreation, gamedesigncocreation2, gamedesigncocreation3}. In this mixed-initiative co-creation \cite{mixed-initiative-cocreativity, mixed-initiativeinterfaces} process, AI acts as a partner of humans and provides real-time feedback aligned with the creation iteration. Since the algorithm can generate numerous instances with easy inputs in a relatively short time, the mixed-initiative interfaces can help its users quickly explore the solution space, inspire them with unexpected ideas \cite{cowriting-aiasactivewriter,co-writing-fromtooltocompanion,cowriting-wordcraft}, and make creative experiences accessible to non-professional creators \cite{co-creation-wideruser-casualcreators,co-creation-wideruser-disables}. 

Current mixed-initiative co-writing interfaces mainly focus on supporting writers. These systems were designed to help writers to keep the consistency of stories, plan plots, get unstuck \cite{unmetmax}, and change text-based stories into other forms \cite{cowritingchangeotherform}. Users must have basic writing skills to operate these systems. Other work introduced gamified designs such as temporary rules and goals \cite{mixedinitiativegame-kuileixi, mixedinitiativegame-wawlt, mixedinitiativegame-looseend, mixedinitativegames-buddy}, as well as scores \cite{mixedinitiativegame-mobile} into mixed-initiative co-writing to make the system more enjoyable to novice writers. However, previous work on human-AI collaboration in the context of creative writing focused on AI as a supporting mechanism to facilitate creative storytelling efforts. Here, I extend prior work by exploring the use of AI for mixed-initiative creative writing as a game mechanic in the context of game design. To design mixed-initiative video games, I aim to explore the following research questions:
(1) What patterns of interaction and player identification emerge in the player-AI co-creating process?
(2) How do game mechanics impact the creation and role perceptions in the process?
(3) How can mix-initiative co-creating be integrated with game mechanics for a unified play experience?

To ground my study, I designed and prototyped \textit{Snake Story}, a mixed-initiative game with the mechanics from “\textit{Snake}” \footnote{https://www.arcade-history.com last accessed 03.06.2023}. The game (referred to as the game version) involved players selecting AI-generated texts or adding their own texts by controlling a growing snake to eat candies on the map, resulting in the creation of a story about a snake. A GPT-3 \cite{gpt3} based language model was employed to provide 2 text selections in each round with different preset parameters. The model would consider previous selections and would write an end for the story when the game or the interaction is over. For comparison, a system (referred to as the non-game version) was also developed for players to directly select AI-generated texts without engaging in gameplay. 

To investigate how players dynamically interact with the game, I conducted a within-subject user study with 11 players (n = 11). Each player was asked to write an approximately 300-word story about a snake in the randomly assigned two versions. Eleven individuals participated in a study where they played Snake Story and their experience was analyzed using a mixed-method approach, including gameplay log data, survey, think-aloud, interview, and observations. Results from the study show: game mechanics significantly affect players' text selection strategies, the quality of the stories, and sense of engagement in creating; players shared a selection strategy for GPT-3 generated texts; different players had different play strategies in both versions and thus perceived themselves and AI differently because of the game mechanics; players with different writing and AI experiences hold different preferences on the two versions. 

In summary, the thesis makes the following contributions to the mixed-initiative game design:
\begin{itemize}
    \item[(1)] I introduce \textit{Snake Story}, a mixed-initiative game for collaborative writing with AI. I present techniques that enable players to write with AI, and I develop both game and non-game interactions and interfaces.
    \item[(2)] In a within-subject study with 11 players, I compared the non-game and the game version and defined:
    \begin{itemize}
        \item [(a)] Players' usage data.
        \item [(b)] Statistic difference between the two versions.
        \item [(c)] Players' strategies for selecting AI-generated texts to create stories
        \item [(d)] Players' play strategies and role perceptions in the two versions.
        \item [(e)] Players' preferences for the two versions.
    \end{itemize}
    
    \item[(3)] Based on the results of the user study, I discuss the design
implications that mixed-initiative games should:
\begin{itemize}
        \item [(a)] Resolve playing and creating conflicts.
        \item [(b)] Increase narrative engagement in playing.
        \item [(c)] Enhance emotional involvement in creating.
        \item [(d)] Balance playing and creating.
        \item [(e)] Find new evaluation criteria.
   \end{itemize}
\end{itemize}

Taken together, these findings guide the design of future engaging co-creation experiences.


\chapter{Related Work}
\label{chap:related work}
\section{Neural Language Models for Text Generation}
Text generation has appeared as a critical application of natural language processing (NLP) technologies, with various applications such as chatbots \cite{chatbotmedical,chatgpttwitter} and content creation \cite{gptcreation}. The rise of deep learning has enabled significant advancements in the field of text generation, with language models such as the Generative Pre-trained Transformer (GPT) series achieving remarkable success. It has been proven that GPT models have the ability to generate texts that cannot be distinguished from human-written pieces \cite{distinguishablegpt2,distinguishablegpt3}. Based on the previous GPT-2 structure \cite{gpt2, attention}, the GPT-3 model with larger model size, dataset size, and more training has demonstrated stronger abilities in text completion and outperformed GPT-2 on several metrics, including fluency, coherence, and relevance \cite{gpt3}. As a result, GPT-3 was employed in the \textit{Snake Story} game. 

By using a few-shot learning approach \cite{gpt3}, the GPT-3 model is able to perform specific tasks under natural language instructions inputted by users. For example, in the \textit{Snake Story} game proposed in this thesis, a prefix "\textit{writing a story of a snake}" was added to restrict the generated texts under the topic. Despite the impressive advancements in text generation, several challenges remain in using GPT-3, including the issue of bias and the difficulty of producing diverse and engaging content. The issue of bias stands for that the generated text may reflect the biases inherent in the training data. Identical prompts in GPT-3 can result in stereotype-related outputs including biases on sex \cite{gpt3bias}, race, and certain religious groups\cite{gpt3bias2}. Also, GPT-3 still has the problem of sometimes generating low-quality texts that repeat themselves, lose coherence over paragraphs, and have contradictory logic \cite{gpt3}. This problem will be enlarged when the parameters of GPT-3 are not set properly \footnote{https://platform.openai.com/docs/api-reference/models last accessed 03.06.2023}.

\section{Mixed-initiative Co-writing Interfaces}

Mixed-initiative interfaces that enable co-creation in various fields have been widely researched \cite{music-cocreation,music-cocreation2,musiccocreation3,drawingcocreation,drawingcocreation2,drawingcocreation3,gamedesigncocreation,gamedesigncocreation2,gamedesigncocreation3}. 
The interfaces can take advantage of exploratory creativity from human writers and the fast generation of diagrammatic lateral paths from generative algorithms to create mixed-initiative co-creativity \cite{mixed-initiative-cocreativity}. Extensive research has explored the potential of mixed-initiative interfaces to aid human writing through editing human-written texts, as well as generating and expanding ideas. Editing and refining functions are the most common functions in the interfaces. For example, Shuming Shi et al. \cite{cowriteedit1} utilized AI technologies to boost users' writing proficiency by enabling them to generate higher-quality text more efficiently. This was accomplished through the implementation of five distinct categories of features in their digital writing assistant, including text completion, error detection, text refinement, keyword-to-sentence conversion (K2S), and cloud-based input methods (cloud IME). Xin Zhao \cite{cowriteedit2} developed a writing assistant that can assist non-native English speakers in overcoming language barriers by offering rewriting alternatives with various tones (such as casual or formal) and lengths (like shortening or expanding).

In collaborative writing, AI can also serve as an idea generator, contributing to the generation of novel concepts and plot lines. For instance, Melissa Roemmele et al. \cite{cowritingideageneration1} created a system that aids users in brainstorming by providing suggestions for the next sentence in a story. Swanson et al. \cite{cowritingideageneration2} described an interactive storytelling system that utilizes a case-based reasoning architecture to offer a range of options for the subsequent sentences, leading the story in entirely diverse directions. Chung et al. \cite{cowritingideagenerating3} introduced an alternative to suggestion-based co-ideation approaches by developing line-sketching interactions that enable users to co-create stories while actively controlling and making sense of the protagonist's fate. Beyond idea generators, AI in \cite{cowriting-aiasactivewriter} was AI was granted a more substantial role as an active writer and assumed responsibility for continuing users' narratives through a unique form of story solitaire. In contrast, Biermann et al. \cite{co-writing-fromtooltocompanion} proposed that AI could jeopardize writers' control, autonomy, and ownership by exceeding co-creative limits, and therefore sought to preserve human control over the writing process in their system. 

Moreover, AI can assist in bridging the gaps between the skeleton structures of stories. Ammanabrolu et al. \cite{cowritingfillgap} introduced an ensemble-based model capable of generating event-driven stories. Yuan et al. \cite{cowriting-wordcraft} built a text editor that can provide plot points that can contextualize the scene built by humans. Laclaustra et al. \cite{cowritingfillgap2} introduced a system that empowered users to specify characters, locations, and objects within a story world. The system, subsequently, generated a rudimentary story by allotting actions to individual characters and creating a sequence of events. 

In summary, the aforementioned applications are well-designed to assist creative writers in enhancing language, ensuring consistency, overcoming writer's block, managing reader experience, as well as refining and iterating on expressive intent \cite{unmetmax}. However, it is crucial to note that more research is indispensable to cater to the needs of casual creators or non-creators in the realm of content creation.



\section{Mixed-initiative Co-writing Games}
\label{section:migforcowriting}

In order to broaden the accessibility of co-creative experiences for a wider range of users, various applications have recognized the benefits of integrating mixed-initiative co-writing as a valuable component of narrative instruments \cite{narrative-instrument} in games, thereby enhancing the overall interactive experience of "\textit{play}".
For example, a mixed-initiative text-based game, \textit{AI dungeon}\footnote{https://play.aidungeon.io/main/home last accessed 03.06.2023}, used AI to generate and respond to players' text-based commands and choices. The AI system produces a distinctive story outcome based on the players' inputs, providing an evolving and personalized gaming experience of exploring and creating stories in pre-set scenes. 
Moreover, Kreminski et al. \cite{mixedinitiativegame-wawlt} developed "\textit{Why Are We Like This?}", a mixed-initiative, co-creative storytelling game that aimed to engage players in investigating the generated history of characters and to bring the story to a satisfying conclusion by selecting and writing actions for the characters. The game involves designed author goals, proper AI suggestions, and player curiosity to encourage co-authorship.

While the term "\textit{play}" is commonly used to denote the interaction between human and mixed-initiative interfaces, it is essential to recognize that "\textit{games}" bear distinctive dissimilarities from play, as they feature unambiguous goals that encourage participants to engage in the interpretation and optimization of rules and tactics \cite{playandgamedefine}. The introduction of goals into a system serves as the most straightforward means of distinguishing mixed-initiative co-writing games from mixed-initiative co-writing interfaces.
Xi et al. \cite{mixedinitiativegame-kuileixi} introduced KuiLeiXi, an open-ended text adventure game that required players to interact with the AI to achieve predetermined plot goals. The system was developed to address the lack of incentives for players in \textit{AI Dungeon}.
Additionally, Ben Samuel et al. \cite{mixedinitativegames-buddy} created a mixed-initiative playful tool, \textit{Writing Buddy}, that integrates the affordances of both authoring and playable media to support creative writing endeavors. The game mechanics prompted players to engage in a puzzle-solving-like experience, where they possessed the freedom to add or eliminate story beats to alter the characters' states within the game and attained the pre-determined narrative goal. 
Building upon the concept of \textit{Writing Buddy}, Kreminski et al. \cite{mixedinitiativegame-looseend} developed \textit{Loose Ends}, a mixed-initiative co-creative storytelling play experience that incorporates obligatory storytelling goals that can be parameterized with specific characters and additional constraints. In addition, the AI system in Loose Ends ensures consistency with all previous texts, which emulates the functionality of an active writing partner. 

The present state of mixed-initiative co-writing games suggests that their full potential has yet to be realized, as they continue to rely on interactions that overlap with mixed-initiative interfaces. While the mobile game designed by Castaño et al. \cite{mixedinitiativegame-mobile} represents a step forward in this field, enabling users to collaboratively create a story by arranging a card-game-like system, further exploration of combining mixed-initiative interfaces and game mechanics are required. 



%

\chapter{Snake Story}
\begin{figure}[htpb]
\begin{minipage}[b]{1.0\linewidth}
  \centering
  \centerline{\includegraphics[width=15cm]{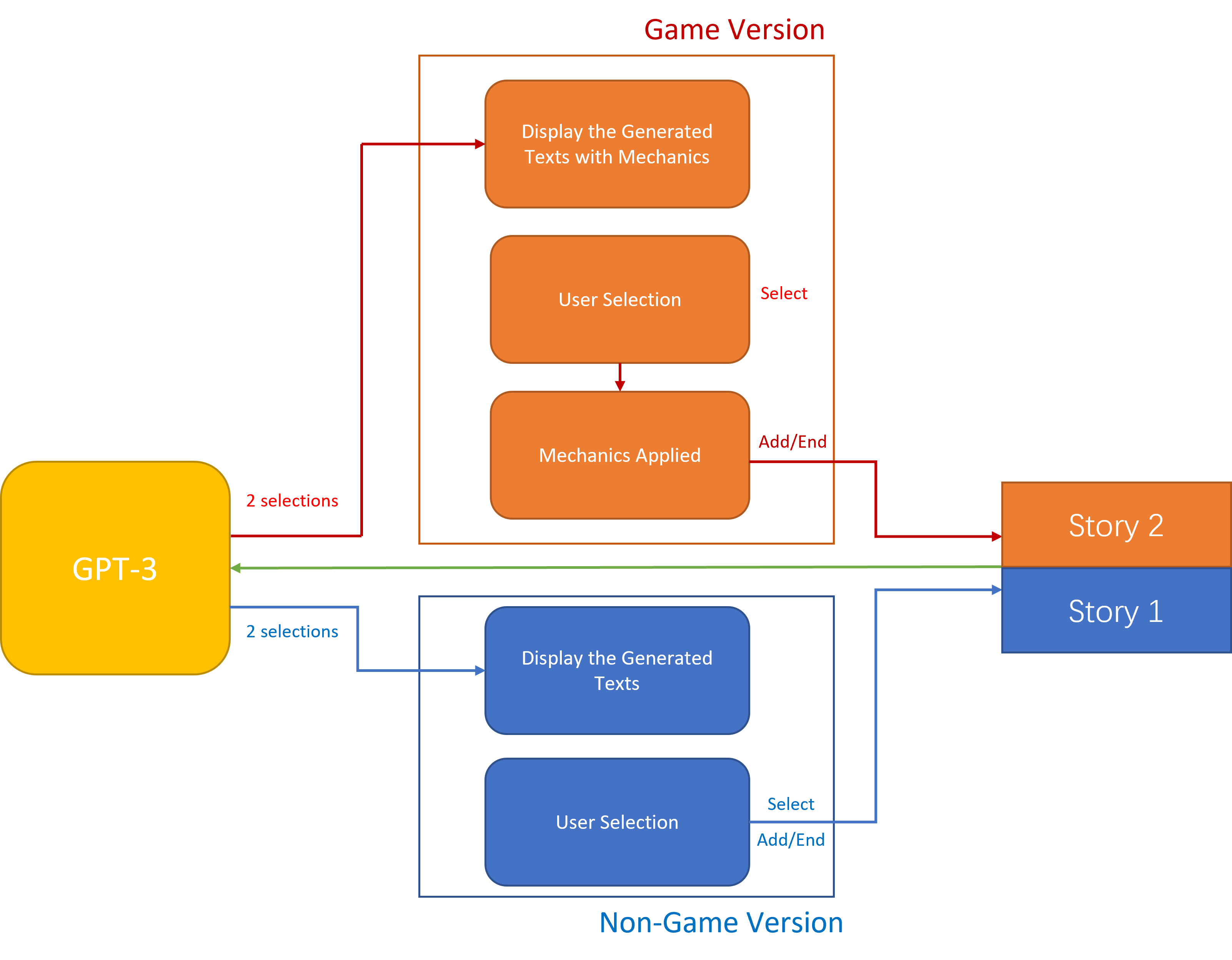}}
\end{minipage}
\caption{The Architecture of Snake Story.}
\label{fig:architecture}
\end{figure}
To ground my study, a mixed-initiative game named "\textit{Snake Story}" was designed and developed in the Unity3D engine. As shown in Fig. \ref{fig:architecture}, the game consists of 2 parts: the non-game version and the game version. The game allows players to write different stories under the same prompt, “writing a story of a snake”, with GPT-3 generated texts in different turn-based interactions. The "text-davinci-003" model \footnote{https://platform.openai.com/docs/models/overview last accessed 03.06.2023} was employed in the system to generate the text.

\label{chap:game}
\section{Non-game Version}

\begin{figure}[htpb]
\begin{minipage}[b]{1.0\linewidth}
  \centering
  \centerline{\includegraphics[width=15cm]{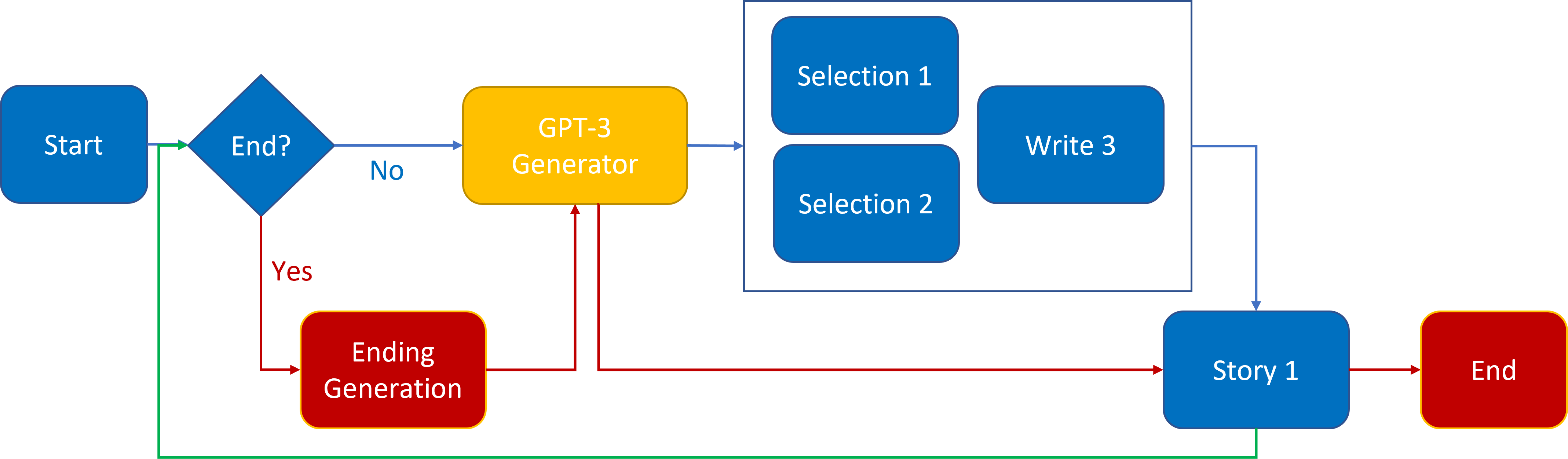}}
\end{minipage}
\caption{The Flow of the Turn-based Interaction in the Non-game Version.}
\label{fig:nongameFlow}
\end{figure}

As illustrated in Fig. \ref{fig:nongameFlow}, the non-game version of the system functions as follows: players are presented with two 30-word text options with different temperatures (0.6 and 1.4) generated by GPT-3 in each turn. If they wish to continue the story, they can select one of the options, which will be automatically added to the narrative. Alternatively, players can opt to compose their own text to continue the story if they are dissatisfied with the AI-generated options. In the subsequent turn, GPT-3 generates two fresh text alternatives for the players to choose from. Once the players decide to end the story, GPT-3 assists in linking the narrative to the predefined ending: ", \textit{and the story of the snake ends}" with a maximum of 80 words.

\begin{figure}[htpb]
\begin{minipage}[b]{1.0\linewidth}
  \centering
  \centerline{\includegraphics[width=15cm]{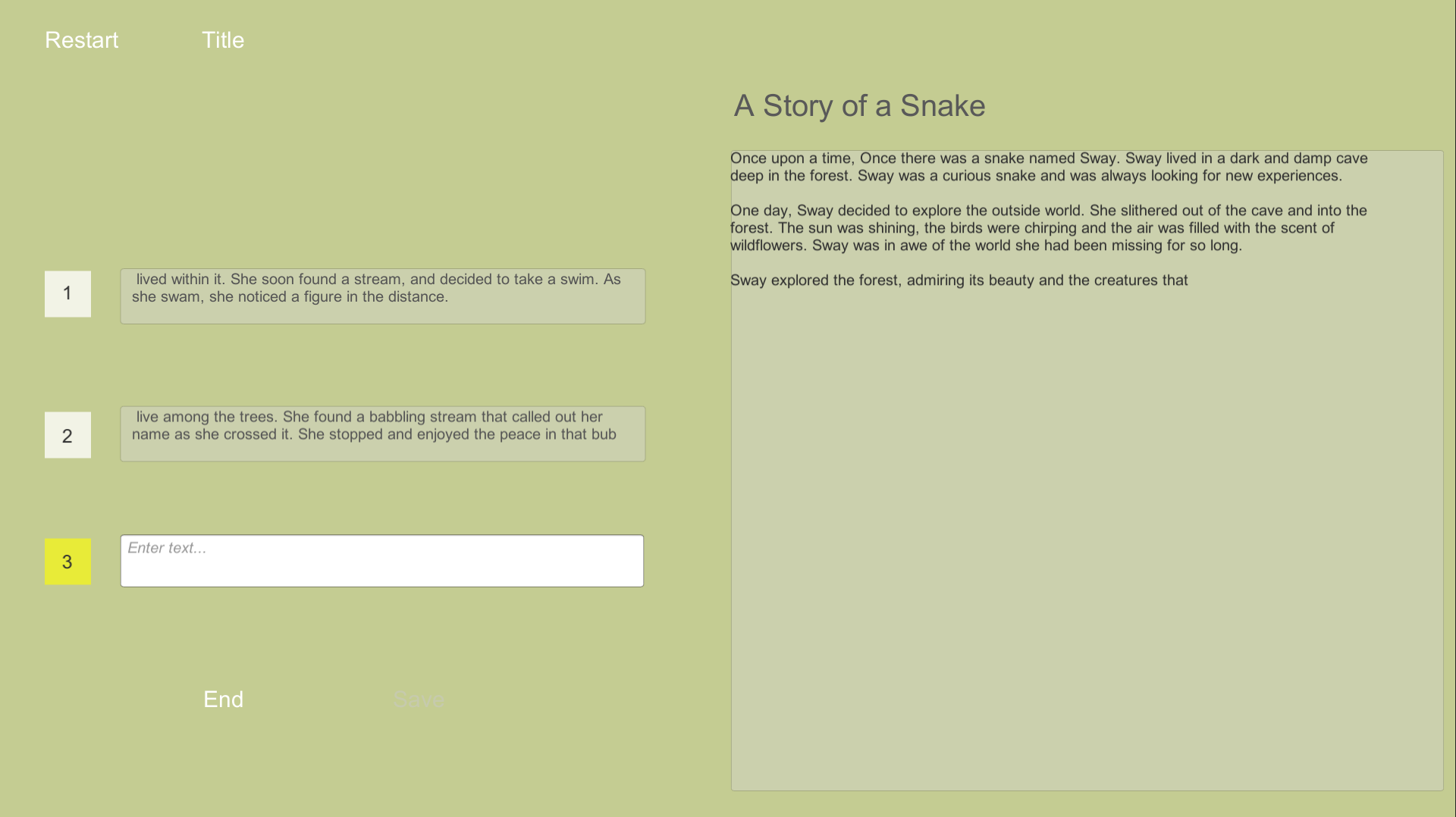}}
\end{minipage}
\caption{The Interface of the Non-game Version.}
\label{fig:nongameinter}
\end{figure}

As depicted in Fig. \ref{fig:nongameinter}, the interface of the non-game version presents the two GPT-3 generated text options on the left side of the screen, accompanied by square buttons containing labels to their left. Additionally, an input field is positioned beneath the text options, enabling players to contribute their own textual content. Once the GPT-3 generation process is completed, the button adjacent to the input field becomes interactable. Players can then click this button to incorporate their selected text into the ongoing narrative, marking the initiation of a new turn. Moreover, an "End" button is situated underneath the text options, providing players with the means to end the story. 

\section{Game Version}
\label{section:game version}

\begin{figure}[htpb]
\begin{minipage}[b]{1.0\linewidth}
  \centering
  \centerline{\includegraphics[width=15cm]{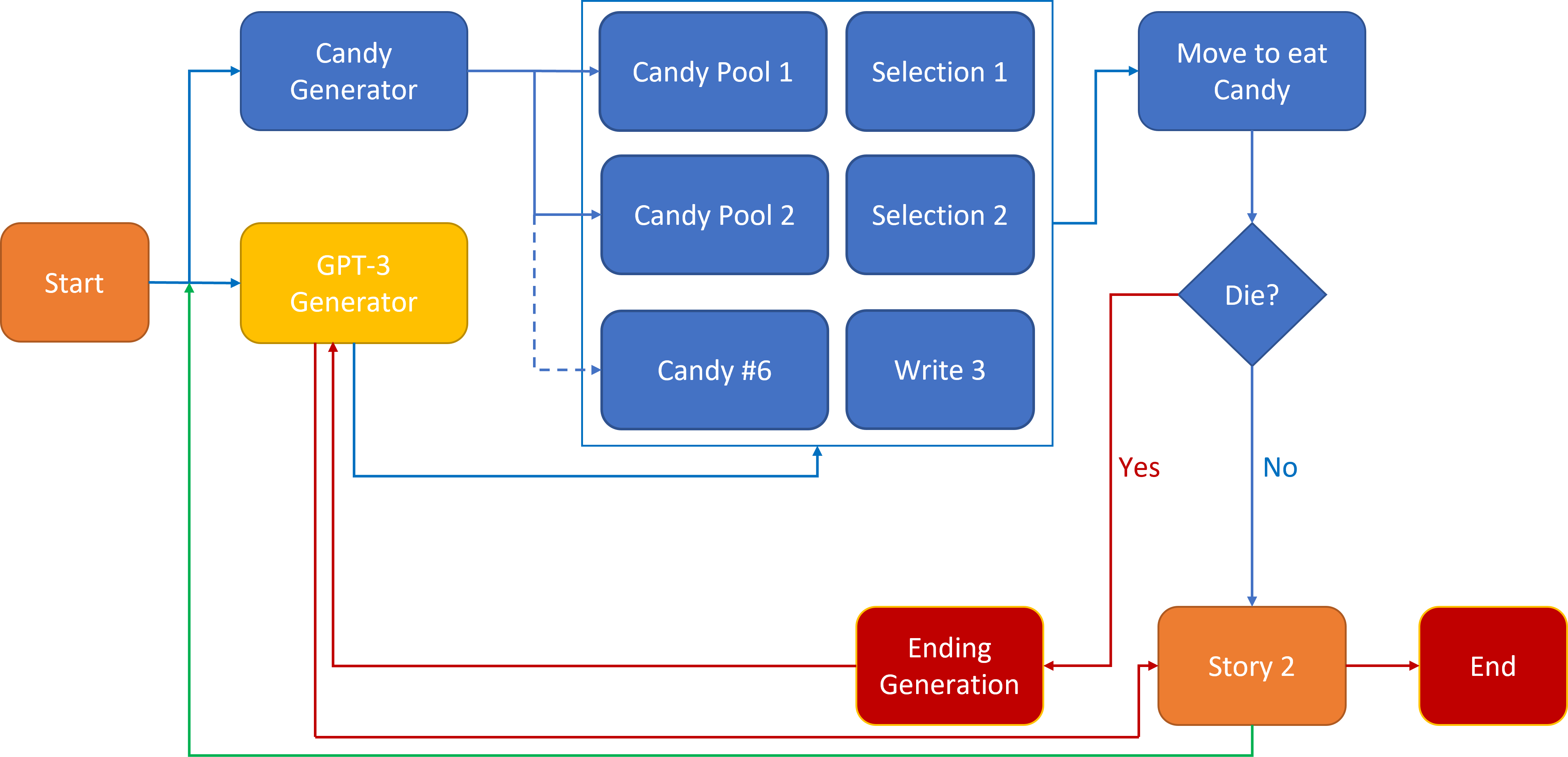}}
\end{minipage}
\caption{The Flow of the Turn-based Interaction in the Game version.}
\label{fig:gameflow}
\end{figure}

In contrast to the non-game version, the game version of the system employs "Snake"-game-like mechanics as a metaphor for adding paragraphs to a story, as demonstrated in Fig. \ref{fig:gameflow}. In the game version, players are still presented with two selections of texts. However, these texts are now represented by candies positioned on a 15*15 tile map, each of which possesses unique mechanics. To add a text to the story, players must navigate a growing snake towards the corresponding candy, which triggers the addition of the selected text to the narrative, along with the application of the corresponding mechanics to either the player or the game map. Players are unable to terminate the story unless their life points become exhausted.

\begin{figure}[htpb]
\begin{minipage}[b]{1.0\linewidth}
  \centering
  \centerline{\includegraphics[width=15cm]{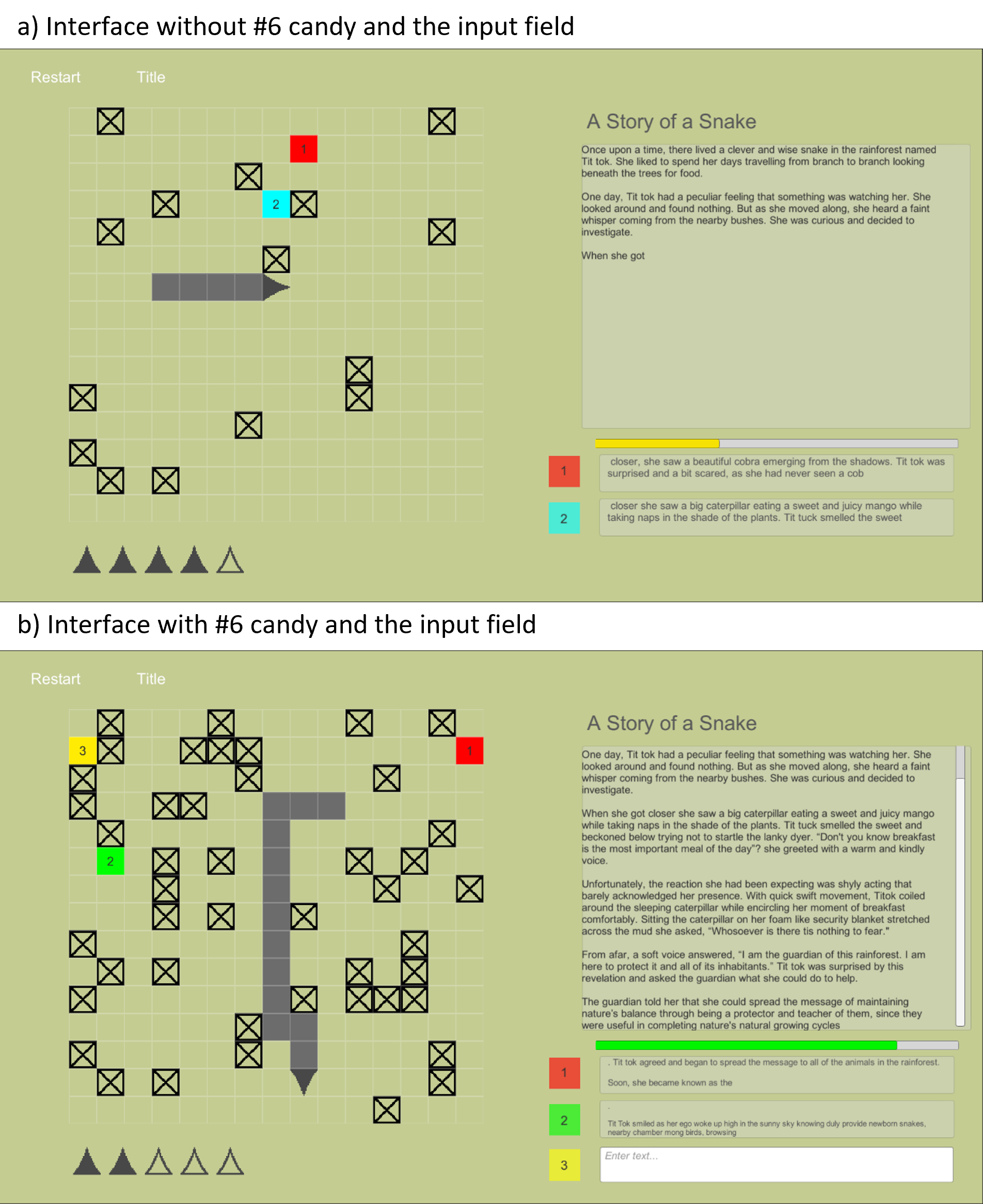}}
\end{minipage}
\caption{The Interface of the Game Version.}
\label{fig:gameinterface}
\end{figure}

As shown in Fig. \ref{fig:gameinterface} a), the game is played on the left-hand side of the screen, while the story is displayed on the right-hand side. Players’ life points are shown on the left bottom under the tile map. The player's life points are located at the bottom-left corner of the tile map. The two text selections, along with their corresponding candies, are displayed under the story. A countdown scrollbar for the pause time is located between the story and text selections, and the game pauses momentarily when new candies and texts appear. Once a player collects the special candy (Blue), they are given the opportunity to contribute to the story by writing their own text. As shown in Fig. \ref{fig:gameinterface} b), an input field will appear under 2 text selections, and a corresponding yellow candy will be generated on the map.

\begin{figure}[htpb]
\begin{minipage}[b]{1.0\linewidth}
  \centering
  \centerline{\includegraphics[width=15cm]{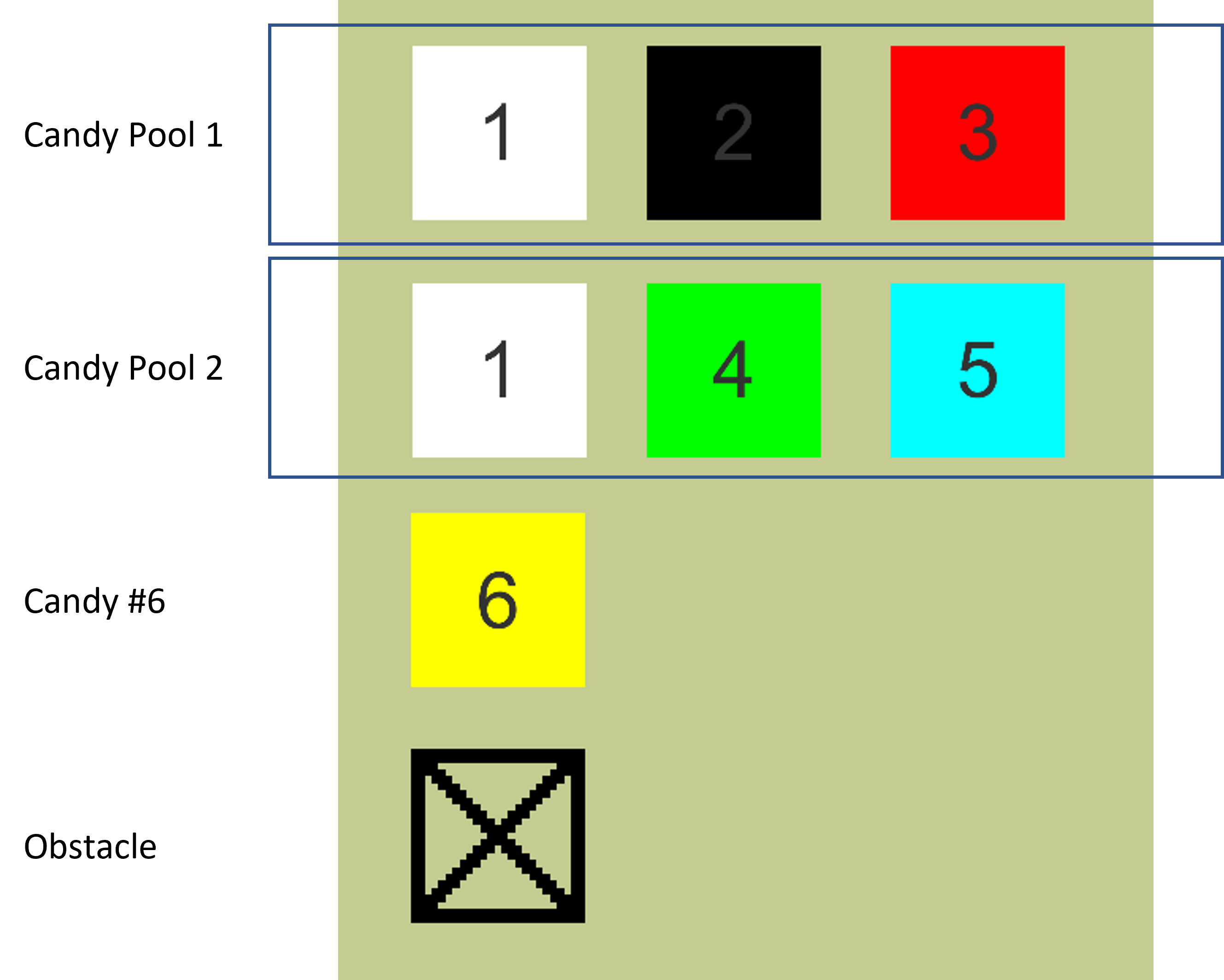}}
\end{minipage}
\caption{Tiles Used in the Game Version.}
\label{fig:candy}
\end{figure}

As shown in Fig. \ref{fig:candy}, seven different tiles are designed in the game, comprising of six types of candies and one obstacle. The candies are divided into two pools: pool 1 for text selection 1 and pool 2 for text selection 2. To research how game mechanics can affect players’ choices of texts, the temperature for selection 1 and candy pool 1 that has negative effects is set lower than that for selection 2 and candy pool 2 for better and more stable text output. Candies with neutral and negative effects are designed for pool 1, which are indicated by negative colors. The white candy, with neutral mechanics, will only increase the snake's length by 1, while the black candy will additionally introduce three extra obstacles on the map. Furthermore, the red candy will decrease the player's life points by 1. Pool 2, on the other hand, features candies with neutral and positive effects as counterparts to the negative candies, indicated by positive colors. The green candy will add 1 life point, while the blue candy will permit players to write their text in the next turn, as demonstrated in Fig. \ref{fig:gameinterface}. After each turn, three obstacles will be added to the map to increase the difficulty level.

In order to investigate the influence of game mechanics on players' text choices, the temperature for selection 1 and candy pool 1 was intentionally set lower (0.6) than that for selection 2 and candy pool 2 (1.4). This decision was made to improve the quality and stability of text output in selection 1. Considering players’ average reading speed and the usage of the think-aloud protocol, the game will be paused for 25 seconds each time when players get new texts. This pause duration will be extended to 45 seconds when players wish to write their own text to add to the story. Players can choose to end the pause early by clicking the buttons adjacent to the text selections, similar to how they would end their turns in the non-game version.

\begin{figure}[htpb]
\begin{minipage}[b]{1.0\linewidth}
  \centering
  \centerline{\includegraphics[width=15cm]{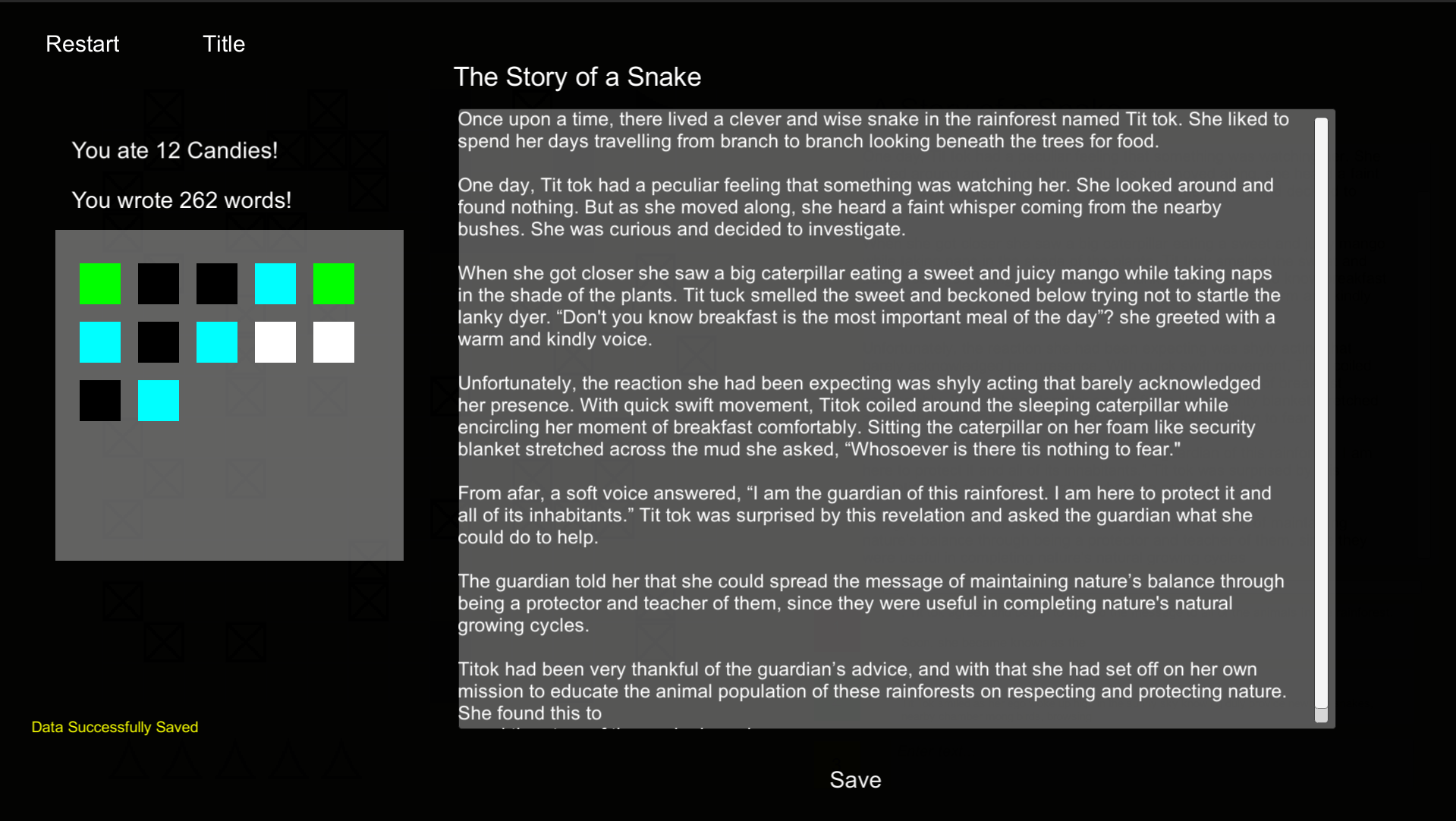}}
\end{minipage}
\caption{The Result Page of the Game Version.}
\label{fig:gamerp}
\end{figure}

When players’ life points become 0, the interaction and the story will end. As shown in Fig. \ref{fig:gamerp}, players will enter a result page.  On the right-hand side of the screen, the full story with an automatically generated ending will be displayed. Additionally, the interface will indicate the length of the snake and story, as well as provide information on the types of candies consumed by the player during gameplay.

\chapter{User Study}
\label{chap:user study}

\section{Participants}

To research how different players interact with \textit{Snake Story}, 11 Game Design students (n=11, referred to as P1-P11) from Northeastern University were recruited through a Discord poster to play the game. Given the premise that the players' writing and AI experience may contribute to their distinct perceptions of the game \cite{co-writing-fromtooltocompanion}, the study recruited a diverse cohort of participants with variable levels of writing proficiency and collaborating experiences with AI. All participants volunteered for the study and were not compensated. 

\section{Procedure}

The study was designed as a within-subject investigation, whereby each participant was assigned to play both the non-game version and the game version of \textit{Snake Story} in random orders. In each session, the participant was given a brief tutorial on the game mechanics and interface and was then instructed to compose a 300-word story about a snake with AI. The participant was also required to engage in think-aloud protocols during the 10-to-15-minute gameplay. Subsequently, the participant was asked to complete a 5-Likert scale usability questionnaire. Following the completion of two sessions, the participants would participate in a semi-structured interview lasting approximately 5-10 minutes, in which they shared their interaction experiences. Finally, participants were asked to complete a demographic survey, which included questions about their writing and AI experience.

\section{Evaluation}

In the game, each text selection generated by GPT-3 was captured and stored. Moreover, the game also recorded the players' selection of texts and the stories they created. To further evaluate the user experience quantitatively, the usability questionnaire incorporated queries on the quality of the generated text, the overall story, and the user's interaction experience.
These collected data were subjected to quantitative analysis, including the use of Wilcoxon signed-rank tests to compare the results from the two versions of \textit{Snake Story}.
During the study, the screen was recorded to capture the participant's interactions with the game, while the audio was recorded to generate transcriptions of the think-aloud protocols and interviews. The resulting data were analyzed using a qualitative approach based on open coding \cite{opencoding}, allowing for a thorough exploration of the participants' experiences and interactions with the game.

\chapter{Quantitative Results}
\label{chap:quant results}

\section{Usage Statistics}


\begin{figure}[htpb]
\begin{minipage}[b]{1.0\linewidth}
  \centering
  \centerline{\includegraphics[width=15cm]{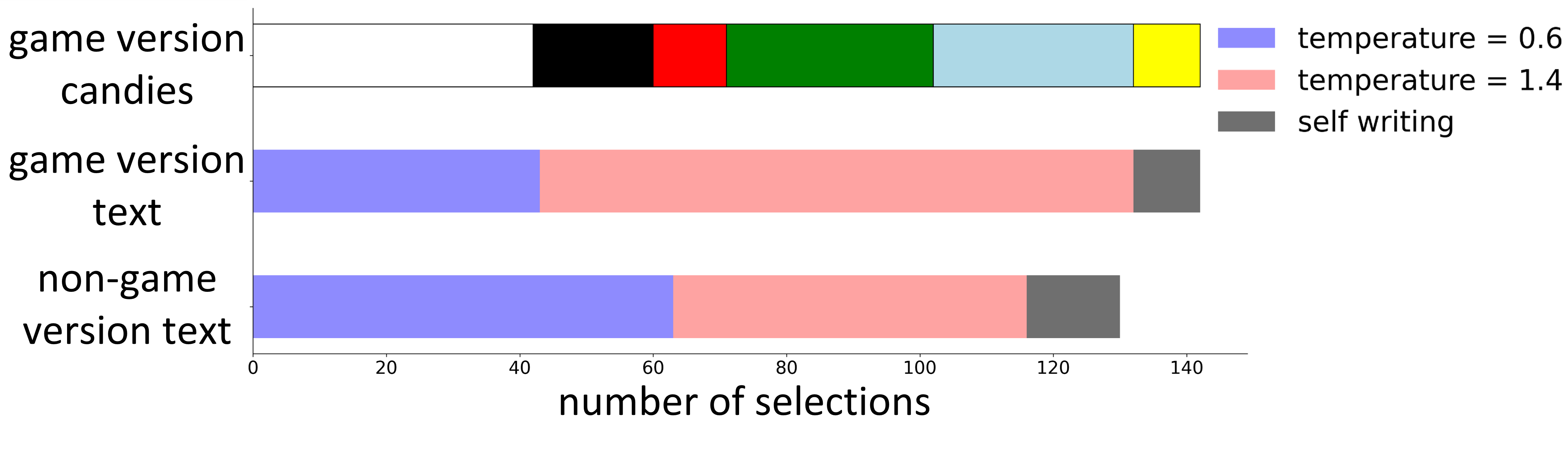}}
\end{minipage}
\caption{Statistics on Players' Text and Candy Selections.\protect\footnotemark[1]}
\label{fig:quan}
\end{figure}
\footnotetext[1]{ Colors in the \textit{game version candies} row match candy colors mentioned in Section \ref{section:game version}}

11 players wrote a total of 22 stories about snakes in 2 versions of the \textit{Snake Story}. The total number for each detailed statistic with an average number (M) and standard deviation (SD) are reported. As shown in Fig. \ref{fig:quan}, the players made a total of 130 choices (M = 11.82, SD = 1.80) in the non-game version. Of these, the generated texts with a lower temperature (0.6) were selected 63 times (M = 5.73, SD = 1.76), while the generated texts with a higher temperature (1.4) were selected 53 times (M = 4.82, SD = 2.48). Additionally, the players chose to write their own words 14 times (M = 1.27, SD = 2.05). On average, the players spent 49.14 seconds (SD = 13.13) making decisions in the non-game version. 

Correspondingly, the players made a total of 142 choices (M = 12.91, SD = 4.50) in the game version. Of these, 0.6 temperature texts were selected 43 times (M = 3.91, SD = 1.98), while 1.4 temperature texts were selected 89 times (M = 8.09, SD = 4.72). Players chose to write their own words 10 times (M = 0.91, SD = 1.83). On average, the players spent 27.33 seconds (SD = 7.69) making decisions in the game version.

In the game, 91 white candies were generated, 42 of which were selected (46.15\%); 50 black candies were generated, 18 of which were selected (36.00\%); 47 red candies were generated, 11 of which were selected (23.40\%); 46 green candies were generated, 31 of which were selected (67.39\%); 47 blue candies were generated, 30 of which were selected (63.83\%); 40 yellow candies were generated, 10 of which were selected (25.00\%).

Wilcoxon signed-rank tests were conducted to compare players' selection and time usage differences in the 2 versions. The test results showed that there was no significant difference in the total number of selections made by players (W(11) = 29.0, p = 0.76). However, the test results showed that game mechanics significantly affected players' choices for 0.6 temperature texts (W(11) = 7.0, p = 0.035). By contrast, it was worth noting that players' choices for 1.4 temperature texts (W(11) = 10.0, p = 0.14) had no statistically significant differences. Moreover, no significant differences were found in self-writing (W(11) = 2.5, p = 0.16) choices between the two versions. Additionally, the analysis indicated that players made decisions significantly faster in the game version (W(11) = 2.0, p = 0.0029).





\section{Story Evaluation}

The stories in the non-game version had an average of 260.64 words (SD = 35.61), while the stories in the game version had 272.64 words(SD = 64.22). There was no significant difference in the length of the stories between the two versions (W(11) = 28, p = 0.70).

Automated writing evaluation tools \cite{awe} were employed to assess the cohesion, grammar, language diversity, and overall writing quality of the 22 stories.
Cohesion was evaluated using two metrics obtained from the Tool for the Automatic Analysis of Cohesion\footnote{https://www.linguisticanalysistools.org/taaco.html last accessed 03.08.2023} (TAAOC) \cite{taaoc}: the sentence overlap rate (S. Overlap) and the paragraph latent semantic overlap rate (P. LSA). 
The Grammar and Mechanics Error Tool\footnote{https://www.linguisticanalysistools.org/gamet.html last accessed 03.08.2023} (GAMET) \cite{gamet} was utilized to detect the number of grammatical errors in the texts. 
In order to assess the language diversity of the writing, the Tool for the Automatic Analysis of Lexical Diversity\footnote{https://www.linguisticanalysistools.org/taaled.html last accessed 03.08.2023} (TAALED) \cite{TAALED} was employed. This tool was chosen for its ability to provide a reliable metric for the measure of textual lexical diversity (MTLD) \cite{mtld}. 
Finally, GPT-3\footnote{https://chat.openai.com/chat last accessed 03.08.2023} itself was used to provide an overall score for the stories on a scale of 0 to 10 \cite{gpt3evaluatstory}.

\begin{table}[htpb]
\centering
\begin{tabular}{||c | c c c c c||} 
 \hline
 \textbf{Metrics} & \textbf{S. Overlap} & \textbf{P. LSA} & \textbf{Error} & \textbf{MTLD} & \textbf{GPT Score}\\ [0.5ex] 
 \hline
 Non-game & 0.19(0.038) & 0.24(0.053)  & 7.00(8.12) & 91.49(65.62) & 6.18(1.90)\\ 
 Game & 0.14(0.053) &  0.20(0.077)  & 10.64(11.20) &159.78(124.18) & 5.63(2.01)\\ \hline\hline

  \textbf{Test Results} & \textbf{Accept}  & \textbf{Reject} & \textbf{Reject} &\textbf{Reject} & \textbf{Reject} \\ \hline
  W(11) & 6.0 &  16.0 & 13.5 & 13.0 & 10.0\\
   p & 0.014 &  0.15  & 0.15 & 0.08  & 0.50\\
 \hline
\end{tabular}
\caption{The Evaluation (M.(SD)) and the Wilcoxon signed-rank tests of AWE.}
\label{table:evaluation}
\end{table}

The results of the evaluations were shown in Table \ref{table:evaluation}. The results from Wilcoxon signed-rank test indicated that the change in text selection preference between versions may impact the cohesion of paragraphs within the stories (W(11) = 6.0, p = 0.014). However, no other significant differences were found in the stories between the two versions.

\begin{figure}[htpb]
\begin{minipage}[b]{1.0\linewidth}
  \centering
  \centerline{\includegraphics[width=15cm]{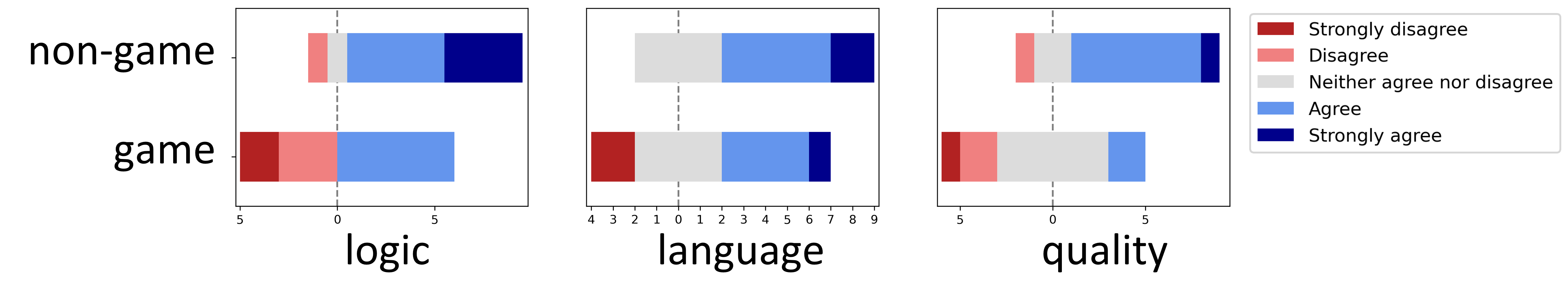}}
\end{minipage}
\caption{Self-evaluations of the Written Stories.}
\label{fig:s1}
\end{figure}

Furthermore, the players were requested to assess the stories they wrote in the 5-Likert scale questionnaire. The results of this evaluation are presented in Figure \ref{fig:s1}. The additional Wilcoxon signed-rank test results indicated that there were no significant differences in the language used in the stories between the game and non-game versions (W(11) = 12.0, p = 0.19). However, the logic of the stories in the game version was significantly weaker than that in the non-game version (W(11) = 0.0, p = 0.0096). Moreover, the overall quality of the stories in the game version was significantly lower than that of the non-game version (W(11) = 3.5, p = 0.020).


\section{Quantitative Experience Report}


\begin{figure}[htpb]
\begin{minipage}[b]{1.0\linewidth}
  \centering
  \centerline{\includegraphics[width=15cm]{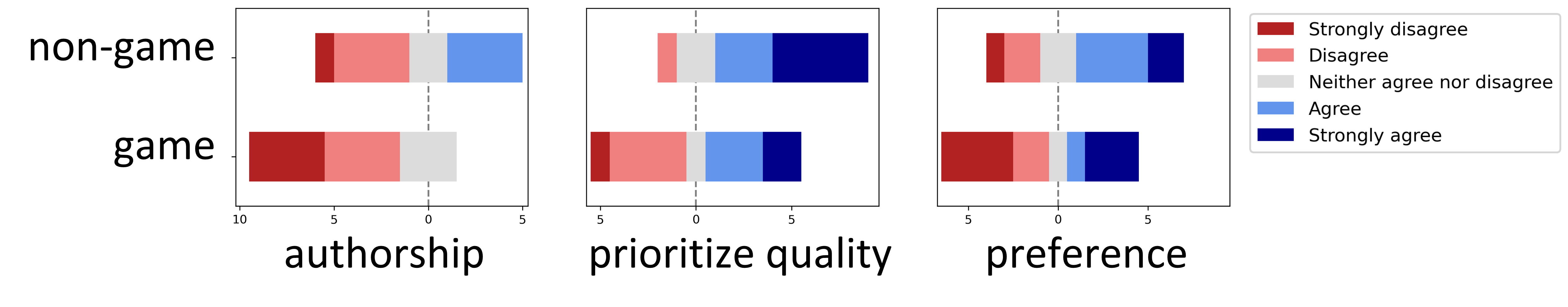}}
\end{minipage}
\caption{Authorship, Creating Strategy, and Preference of the Written Stories.}
\label{fig:s2}
\end{figure}

As shown in Fig. \ref{fig:s2}, through the implementation of Wilcoxon signed-rank tests on the questionnaire data, it was observed that players had significantly less authorship of the story in the game version (W(11) = 3.0, p = 0.031). Furthermore, the players showed a significant difference in their interaction goal between the two versions, as they placed a greater emphasis on prioritizing the quality of the stories in the non-game version (W(11) = 2.0, p = 0.040). Nonetheless, no significant statistical difference was detected in their preference for the stories across the two versions (W(11) = 4.0, p = 0.083).

\begin{figure}[htpb]
\begin{minipage}[b]{1.0\linewidth}
  \centering
  \centerline{\includegraphics[width=15cm]{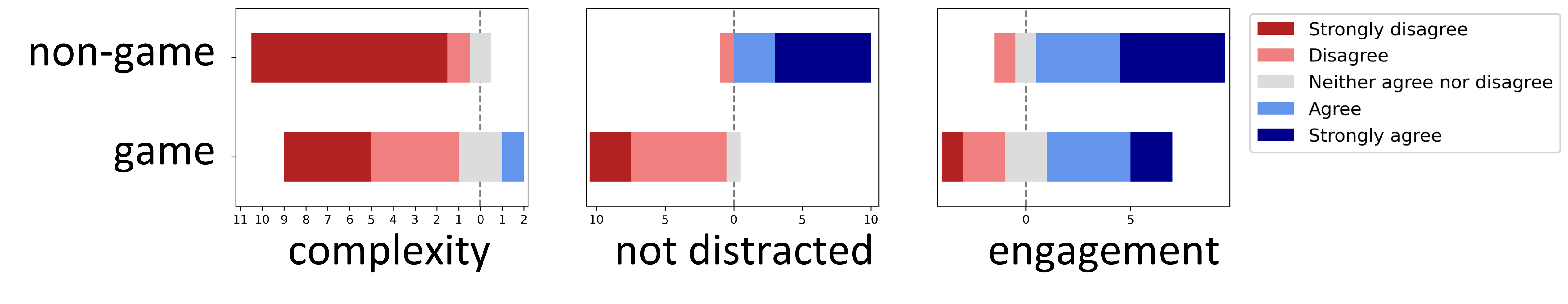}}
\end{minipage}
\caption{The Complexity, Distraction, and Engagement of the Two Versions.}
\label{fig:s3}
\end{figure}

Additionally, players rated their interaction experience in the questionnaire. As shown in Fig. \ref{fig:s3}, players had significantly different perceptions between the two versions. Specifically, The game version was perceived to be significantly more complex compared to the non-game version (W(11) = 0.0, p = 0.039). Additionally, interactions within the game version were reported to have a significant impact on the creation process (W(11) = 0.0, p = 0.00098), whereas the creation process in the game version was considered to be less engaging (W(11) = 2.5, p = 0.047).


\chapter{Qualitative Results}
\label{chap:quali results}

\section{Interaction Patterns in the Non-game Version}

\subsection{Text Selection Strategies}
\label{subsection:nongame text strategies}
After analyzing the think-aloud data of players in the non-game version, 124 open codes were identified into 5 distinct categories based on the explanations the players provided for their choices. These categories are language (24), consistency (69), unexpectedness (17), self-writing reasons (12), and other situation (2). 
\subsubsection{Language}

Players tended to choose texts of higher language quality, particularly those containing detailed descriptions, elaborate adjectives, and emotional expressions (19). For example, P9 mentioned "\textit{Although both 2 texts were cohesive to the previous texts, the description of snake behavior in Text 1 is more specific.}" when selecting between "\textit{...to explore the inside, and soon found himself submerged in the knee-depth water. He made his way from lily pad...}" and "\textit{...and he quickly jumped into the water. He swam around for hours, enjoying the cool and refreshing sensation of the pond's waters. As...}".

Additionally, players demonstrated a preference for texts that were well-structured and composed with a professional tone (5). For instance, P3 mentioned "\textit{I think the 2 selections are similar, but the second one is more professional and I will go with this one.}" when selecting between "\textit{...He had seen many creatures come and go in his long life, but he was content with his own company. He kept to himself, and the...}" and "\textit{...Named Anaconda, known by every passing creature in pursuit of warmth. One could often hear laughter ringing near it’s solace...}".
 
\subsubsection{Consistency}

Players preferred the texts that aligned with their pre-determined tone (24). As an illustration, P5 pointed out "\textit{I would select 1 because 1 is more like a start of a fairy tale. I do not want to write a realistic story so I will choose 1.}" when selecting between "\textit{Once upon a time, there was a small snake who lived in the forest. She was very curious and loved to explore her surroundings. One day...}" and "\textit{Once there was a green, spotted snake who mind made his home in the deep parts of lush tropical jungle. This snake was quite different than other...}".

Moreover, players demonstrated a proclivity towards selecting texts that unfolded the story in their anticipation (15). Such as P2 stated "\textit{I wanted to put my snake in a safe situation, (because) I don't want my snake to die. (Choose 1)}" when choosing from "\textit{...-green scales glinted in the sun. Alice was sure that this snake wasn't dangerous, and she certainly didn't want to...}" and "\textit{...shadowed a lower tear of its cheekbones. 'Hello there,' She eagerly greeted the glowing-eyed serpent and for just a few...}".

Furthermore, players exhibited an inclination towards texts that maintained coherence with preceding texts, specifically those that exhibited sound and expandable logic (30). As an instance, P7 said "\textit{I think a snake cannot wear the necklace. Also, the golden necklace is so luxurious that it does not seem like something a bird that has just been saved would have.}" when selecting between "\textit{...special gift. It was a beautiful golden necklace with a single ruby in the center. Slither was amazed by the gift and decided to wear it around...}" and "\textit{...(a)n Acorn sap, that would grant Slither fortune to any human being wished aide of him. Slither couldn't wait to tell the...}".

\subsubsection{Unexpectedness}    

Players displayed a preference to select unexpected texts that featured fresh settings, new characters, and surprising plot twists (11). To illustrate, P3 explained "\textit{I think 1 is more fun. It has new people (objects). 2 just mentioned familiar faces. I don't like familiar faces}" when choosing between "\textit{...it was surprised to find a new world of exotic animals, plants and trees. It found itself in an oasis full of life and beauty,...}" and "\textit{...it met the familiar faces, who watched without any hesitation. The explorative beast evolved into steps of understanding slowly and carefully forging relationships while exchange...}".

In addition, players showed a propensity to select texts that possessed a sense of suspension regarding their potential narrative developments (6). For example, P11 said "\textit{I want to see where this goes. I chose this (2) because it's messed up, and I want to see if it becomes more messed up if I choose the messed up option.}" when selecting between "\textit{...I grant ye the power to control the elements of this world, but only when you accept my blessing.' George was terrified and uncertain what...}" and "\textit{...Gallivanting about every daye frivolipaseday joorneys with on larkining flightal skeemeshyne lizard wingable sprites...}".

\subsubsection{Self-writing Reasons}

In the situation that neither of the presented text options fulfilled their preferences, players were observed to write their own content, which frequently drew inspiration from the provided text selections (6). As an example, P10 mentioned "\textit{The first one is cheap...I like the 'anger' part (in 2), but then I don't like the 'mouth puckering', so maybe I can do that...}" when facing "\textit{...and eventually let out a soft purr before turning away and walking off into the distance. The snake was relieved that he had been spared,...}" and "\textit{...and anger never leaving his eyes. He narrowed his focus to the powerless serpent before him, is mouth puckering upward ready to end this mercy mission of...}", but finally writing "\textit{..., but anger never leaving his eyes. It calmed down eventually and let out a soft purr before turning away and walking off into the distance. The tiny snake was relieved that he had been spared,...}"

Players' desire to play with the AI was another factor that motivated them to write their own content (6), as will be illustrated in Section \ref{subsection:nongame play strategies}.

\subsubsection{Other Situation}

In a rare situation, players indicated satisfaction with both text options and selected one randomly (2). P7 said "\textit{I think both of the selections were good and appealing to me. Can I randomly choose one of them? (After performing a random selection procedure,) OK, I will go with 1.}" when choosing between "\textit{...other animals of his luck, but first he wanted to test the sap. He poured some onto a nearby rock and wished for more food. Suddenly,...}" and "\textit{...animals in the Kindgdom about this wonderful gift! He spread word quickly, and sure enough many of his animal friends began asking for his help....}".


\subsection{Role Perceptions and Play Strategies}
\label{subsection:nongame play strategies}
Five players use AI as a writing assistant (WA) to support their writing. Three of these players, who identify themselves as writers, believed that they had made the majority of the contributions to the story. For example, P5 said "\textit{Well, even though the AI generated most of the content, I still feel like I had a significant role in creating the story because I made the choices on how the AI wrote. So, I believe that I can claim authorship of this story.}" in the interview. The other two players claim less authorship of the story and describe themselves as "puppet masters" according to P3. P6 also shares this sentiment, stating '\textit{I think I am just providing prompts, and the AI can help me to link them.}'.

Four players consider AI to be an active writer (AW) that provides stories for them to read. They would describe themselves as readers of an interactive storybook, where the AI is the author, and they are the audience. For instance, P10 mentioned "\textit{I'm not planning on writing much on my own. I'm actually more interested in seeing what the AI comes up with.}" before starting to play the non-game version.

The two remaining players consider AI as a playful tool (PT) and engage in challenging or tricking the AI by actively using self-writing functions to generate unexpected or amusing outcomes. They view AI-generated texts not only as a means of co-creation but also as a source of entertainment, exploring the limits and capabilities of the system. To illustrate, P6 mentioned "\textit{I think AI is pretty good at generating texts based on cohesive inputs, but I'm curious to see how it can handle unexpected situations. So I'm gonna test it out by seeing what happens if I just kill off the snake in the story and let the AI continue from there.}" when adding the sentence "\textit{The snake is dead.}" at the very beginning of the story.

\section{Interaction Patterns in the Game Version}

\subsection{The Effect of Mechanics}
\label{section:effectofmechanics}

All players acknowledged that the mechanics had a significant impact on their co-writing process. 
The overall game design, particularly the time limit mechanics, had a significant influence on how the players read the generated texts. Two out of eleven players reported that they never read the generated text. For instance, P5 mentioned that "\textit{Given that it's a game, I'm not particularly concerned about what the AI writes. My main focus is on the gameplay itself.}" By contrast, four players read the text in its entirety, but only intermittently as they controlled the snake to avoid obstacles. For example, during the 5th round, P11 commented, "\textit{I think I can find a safe path for my snake to stay in, and then I can have extra time for reading the texts. Oh, this works!}". The remaining five players opted to give the generated texts a quick scan. To illustrate, P8 mentioned "\textit{So basically, I just skim through the text real quick cause I also need to focus on figuring out how to get my snake to chow down on what I picked out for it at the same time.}" in the interview.

Additionally, the candy mechanics influenced players' choice strategies. Despite their low-quality text, the green and blue candies (good candies) are particularly attractive to players.
To illustrate, P2 mentioned "\textit{I would more likely go for the green candy to regain my lost HP and keep myself alive in the game for a bit longer.}" while playing the game. By contrast, black and red candies (bad candies) are rarely chosen by players. For example, P7 mentioned that "\textit{Even though the texts in the black candy are better, I'm not really keen on making the game more challenging. Plus, the white candy's text is good enough for me.}" However, in situations where white candies are present alongside white or "good" candies, or when a player's health points are at a safe level, they are more likely to apply their selection strategies from the non-game version for text content. Such as P11 said "\textit{The (black) one I chose just now was the more sad option, and I am choosing to make this snake's life sad.}" selecting between \textit{"...The snake was quite content with this lifestyle until one day, when it heard a strange noise coming from the house. Curiosity got the better...}" (black) and "\textit{...Seasons unearthed happiness all around this cozy old home, kids screeched and teenage gossip shook the foundation undoubtedly our friend welcomed shelter in...}" (blue).

Furthermore, the intentional design of the obstacles in the game resulted in a notable increase in emotional arousal among players during the co-creation process. Players were found to experience a range of negative emotions, such as tension and frustration, when attempting to navigate and avoid obstacles, or when inadvertently colliding with them. 


\subsection{Role Perceptions and Play Strategies}

Despite all players identifying as "players" in the game version, their respective play strategies exhibited significant variation.
The majority of the players (7) will make trade-offs (T) between game mechanics and writing. These players aimed to uphold the story's integrity but were willing to compromise its quality if it meant prolonging the snake's lifespan in the game. As an illustration, P11 mentioned "\textit{I'm really tempted to pick the red option, but I know it'll end up killing me, so I'm gonna have the other one. I'd rather keep myself alive in the game to see more stories.}" 

However, four of the players ignore (I) the writing systems and just merely focus on the "\textit{Snake}" game. These players indulge in either the good or bad candies exclusively during gameplay, purely maintaining the life of the snake or increasing the difficulty of the game for fun. For example, P1 stated "\textit{Even I want to choose the texts but the mechanics keep me away. To be honest, I'd much rather focus on enjoying the gameplay rather than putting effort into crafting a compelling narrative.}" in the interview.


\section{Preference}
\label{section:Preference}
\begin{table}[htpb]
\centering
\begin{tabular}{||c | c c c c c c c c c c c||} 
 \hline
 \textbf{Player Codes} & \textbf{P1} & \textbf{P2} & \textbf{P3} & \textbf{P4} & \textbf{P5}& \textbf{P6}& \textbf{P7}& \textbf{P8}& \textbf{P9}& \textbf{P10}& \textbf{P11}\\ [0.5ex] 
 \hline
\textbf{AI experiences} & N & Y & Y & N & N & Y & N & Y & N & Y & N\\ 
\textbf{Writing experiences} & R & P & N & R & R & P & N & R & N & N & N \\ 
\textbf{Non-game V. RP}  & WA & WA & WA & WA & WA & PT & AW & AW & AW & AW & PT \\ 
\textbf{Game V. PS} & I & T & I & I & I & T & T & T & T & T & T \\ 
\textbf{Preference} & NG & - & - & - & - & NG & G & NG & G & - & G \\ \hline
  
\end{tabular}
\caption{The Preference Table of 11 Players.\protect\footnotemark[1]}
\label{table:preference}
\end{table}
\footnotetext[1]{\textbf{AI experiences}: N (No), Y (Yes); \textbf{Writing experiences}: R (rich), P (Poor), N (NO); \textbf{Non-game Version (V.) Role Perception (RP)}: WA(Writing Assistant), AW (Active Writer), PT (Playful Tool); \textbf{Game Version (V.) Play Strategies (PS)}: I (Ignore Writing), T (Make trade-offs); \textbf{Preference}: NG (Non-game Version), G (Game Version), - (No preference)}

As shown in Table \ref{table:preference}, most of the players (5) did not demonstrate a discernible inclination towards either the non-game version or the game version. While they believed that the non-game version was more suitable for serious writing, they found the game version to be more entertaining and enjoyable. For example, P5 mentioned "\textit{If I wanna write a story, I would choose the first one (the non-game version). But for fun, I would play the second one (the game version).}" in the interview.

However, three players (P1, P6, and P8) expressed their strong dislike for the game version, stating that it significantly impaired their creation and reading process. For instance, P8 explained "\textit{I didn't think it was as fun as the other version of the game. I thought it was a little stressful ... if you enjoy that type of narrative (reading or writing a story as it unfolds), I think the first one is, gonna be more appealing.}" in the interview.

 Nevertheless, the remaining three players (P7, P8, and P11), who had neither AI nor writing experience, expressed their strong admiration for the game version. They believed that the challenges presented in the game version increased their engagement in the creation process. As an illustration, P9 stated "\textit{I like the game version more. I think the challenge in the game makes me more engaged in the interaction. The sense of tension in the game version makes it harder for me to consider each selection thoroughly. This means I'm always looking forward to the next choice, hoping to make better decisions than before.}" in the interview.


\chapter{Discussion}
\label{chap:Discussion}

\section{Resolve Playing and Creating Conflicts}

Designing mixed-initiative games with consideration for the potential conflicts between gameplay and creative content generation is essential to promote engagement in the co-creating process. Mechanics that allow for both play and creativity to coexist can encourage players to develop their own unique stories and experiences within the game world. Specifically, as discussed in Section \ref{section:migforcowriting} and Section \ref{section:effectofmechanics}, clear rules and mechanics in \textit{Snake Story} can pose additional challenges for players who wish to engage in creative content generation, particularly when their writing goals (write a better story) conflict with the playing goals of the game (live longer). To mitigate such conflict, exchanging the temperature between good and bad candies can incentivize players to focus on both keeping the snake alive and generating high-quality stories. However, it is important to note that some intrinsic conflicts between playing and creating cannot be easily resolved through such parameter adjustments. In such cases, more specialized and deliberate mechanics must be designed. For example, \textit{Snake Story} has an emergent endpoint when players run out of life points, whereas players' stories may continue, making it difficult to determine a definitive end for them. One possible solution for this issue can be a Neo-rogue-like game system with permanent death mechanics \cite{neorogue} that enables players to continue creating a larger story despite dying multiple times.


\section{Increase Narrative Engagement in Playing}

Developing a tight narrative link between game mechanics and co-created content is a crucial factor in augmenting the participants' sense of immersion in mixed-initiative games. Although \textit{Snake Story} was designed based on a metaphorical representation of the manipulated snake as the snake in the story, a majority of the players (n=7) expressed their dissatisfaction with the perceived disconnection between the game and the narrative. 

Two possible directions can be applied to \textit{Snake Story} as well as future mixed-initiative games. The first direction involves simulating the creative process of renowned writers, such as Shakespeare, in crafting a story. This would involve modeling how such writers generate and develop various ideas, unfold plots, and navigate potential challenges in their writing process. In the game, AI would be leveraged to simulate the thought processes of these writers \cite{discussionshakespear}, while game mechanics can enable players to actively participate in the co-creation of the story by engaging in this abstract thinking process.

Alternatively, players can be cast as the main character of the co-created story. This can be accomplished through an interactive drama game design \cite{interactivedrama,discussion1001}, wherein players take on the role of the protagonist and make consequential decisions that influence the story's direction. To enhance player immersion and emotional investment in the story, personalized elements reflecting the player's experiences and characteristics can be integrated using AI. However, since players' interests align with those of the characters, conflicts between playing and creating must be resolved through additional mechanics.




\section{Enhance Emotional Involvement in Creating}

To mitigate player frustration, mixed-initiative games should incorporate a degree of flexibility that allows players to manage unforeseen emergent events that may arise during gameplay or the creative process. For instance, in \textit{Snake Story}, as discussed in Section \ref{section:effectofmechanics}, players experienced frustration when they were unable to allocate sufficient time to planning the story and maneuvering the snake simultaneously. To address these concerns, a mechanic could be incorporated that enables players to conserve unused time during easier situations and then utilize it during more challenging scenarios. This flexible design can decrease player frustration by introducing a feeling of control, while still retaining the intensity of the gameplay experience.

Moreover, given that mechanics have the potential to exert a noteworthy influence on players' co-creation strategies, mixed-initiative games can employ incentivization through game mechanics as a means of fostering engagement in the co-writing process. For example, in the \textit{Snake Story}, favorable outcomes can be associated with the acquisition of the yellow candy, thereby stimulating players to generate their own textual content.





\section{Balance Playing and Creating}

Similar to the significance of traditional games keeping players in an optimal state of flow \cite{flowingames}, mixed-initiative games should maintain a good balance between playing and creating for players. There are different positive feedback mechanisms between gaming and creative endeavors. Gaming requires short-term, rapid feedback, while creative endeavors often involve long-term, slow feedback. As mixed-initiative games require the players to both engage with game mechanics and creative content generation, it is crucial that the game design facilitates a smooth transition between these two modes in its gameplay. This can be achieved through thoughtful design of factors such as game pacing and player agency. 
Furthermore, a well-designed mixed-initiative game should provide players with appropriate guidance and tools to enable them to create meaningful and enjoyable content, without feeling overwhelmed by the creative demands of the game.

In addition, it is imperative to account for individual differences when designing mixed-initiative games. This is due to the fact discussed in Section \ref{section:Preference} that varying players may necessitate distinct interaction strategies, thereby necessitating a tailored approach to maintain optimal playing-creating flow. Additionally, AI should consider the unique creating strategies (as described in \ref{subsection:nongame text strategies}) of each player to generate personalized content that aligns with their writing goals. The integration of player-centric AI content generation can help to keep players in the flow by reducing low-quality options and providing uplifting text at the appropriate time. 
%

\section{Find New Evaluation Criteria}

To achieve a unified experience of creating and playing in mixed-initiative games, it is crucial to establish novel evaluation criteria that can fairly assess players' creative behavior. This is because an unfair assessment may lead to player frustration and undermine the gameplay experience. While the use of automatic writing evaluation \cite{awe} was demonstrated in the study as a post-evaluation method for the stories, its applicability to evaluating writing quality within the game may be limited by its statistical nature, which might not be applicable for an individual's writing and does not consider subjective player perceptions. Furthermore, real-time human evaluation is not a feasible option. As such, a potential solution could involve the development of a novel algorithm to evaluate players' work automatically. Alternatively, a better approach could involve incorporating game mechanics that allow players to self-evaluate or rate each other. However, the effectiveness and feasibility of these approaches need further investigation.

Additionally, while current evaluation criteria for traditional games may still apply to some extent, mixed-initiative games involve unique features and require new criteria to accurately measure their effectiveness. Mixed-initiative games require new evaluation criteria that account for both the game mechanics and the effectiveness of the mixed-initiative interface. Specifically, it is important to assess how the game mechanics are dynamically combined with the mixed-initiative interface. Nevertheless, the evaluation of mixed-initiative games is still an area that requires further research to establish effective criteria and methodologies



\chapter{Conclusion}
\label{chap:conclude}

In conclusion, the paper presents a prototype of a mixed-initiative game, \textit{Snake Story}, aimed at exploring gamification of co-creation interactions between humans and AI. The study involved 11 participants, who were asked to play two versions of the game with and without game mechanics. 
The finding suggested that mechanics might significantly influence the players' creative processes and self-identifications in the game. Additionally, players with different backgrounds showed different preferences for the two versions. Overall, the study highlights the potential of gamification in making human-AI interactions accessible and fun for players and provides valuable insights for the design of engaging co-creation experiences.

\bibliographystyle{plain}

\bibliography{bib/thesis}

\newpage
\appendix

\chapter{Stories}
\section{Sample Story 1 (Game Version P8)}

Once upon a time, there was a snake who lived in the woods. He had never seen another creature like himself, but he enjoyed living among the trees and eating the small animals that lived there. 

One day, he noticed something strange in the distance - a long, winding line of creatures like himself! He decided to investigate, and as he got closer, he realized that the line was a group of snakes making their way through the forest. 

He was so excited to see other snakes like himself that he quickly joined the procession, slithering alongside them as they moved through the forest. Along the way, he felt particularly drawn to one snake in particular, who would often look back and pause for just a moment as if to acknowledge his presence. 

After a few days, the procession reached its destination - a beautiful, secluded lake in the middle of the forest. The snakes quickly dispersed, but the snake that he had been drawn to stayed behind and waited for him. 

The two of them shared a moment above the waters as they looked into each other’s eyes, acknowledging their instant connection. From then on, it was only the two of them and their limitless adventures among the trees and by the lake, both content to live life together in this idyllic home. 

\section{Sample Story 2 (Non-game Version P8)}

Once upon a time, there was a small snake named Lucy. She lived in the woods near a small village and often ventured out during the night when things were still and quiet. Every day was the same for Lucy, scampering among the earthy loam of the forest floor in search of insects and grubs to satisfy her hunger. 

But one night on her usual midnight march, something stopped Lucy in her tracks – a basket of fruits, vegetables and other goodies had been left outside the village gates. Lucy was curious and hungry so she slithered closer to investigate. As she inched closer, Lucy noticed that the basket was guarded by a large and intimidating snake. He had a long body with shimmering golden scales and a sharp, pointed tail. Lucy knew that this was no ordinary snake – it was a cobra!

The cobra noticed Lucy and coiled itself around the basket as to challenge her. Even with her tiny size, Lucy stood up and faced off against the cobra. Still her bravery paid off and the cobra slithered away, allowing Lucy to feast on all the goodies inside. 

From that day forward, Lucy became known as the brave little snake who stood up against a cobra. She was respected and admired by all of her forest friends, and even the villagers began to leave treats outside the gates for her. Lucy lived a long and happy life in the woods, always remembered as the brave and intrepid, little snake.

\section{Sample Story 3 (Game Version P9)}

Once, there lived a majestic green snake in the heart of a untouched forest. Its piercing fire suffused its emerald body as it knowingly crawled through the foliage.

The snake had a special affinity for humans and often followed them around their camps, watching from afar as they cooked, talked, and laughed. It had no fear of them and often interacted with them in a friendly manner - though some people were scared of it because of its size.

One day, the snake was out exploring a new part of the forest when it stumbled across a mysterious stone altar with strange symbols carved into it. It was intrigued and decided to investigate further, only to find that the altar held a powerful magical gem.

The snake quickly realized that the gem had the power to grant wishes, and it began to think of all the things that it could wish for. After much deliberation, it decided that it wanted to fly so that it could see the world beyond its forest home. So, with a passionate final wish, the snake found itself rising into the air and soaring through the sky.

It was a liberating experience for the snake, and it enjoyed every second of its newfound freedom. From that day forward, the snake was able to explore distant lands and experience new cultures. It even made friends with other animals along its journey.

The snake was truly happy, and it would never forget the day it found that magical gem.

\section{Sample Story 4 (Non-game Version P9)}
Once upon a time, in a grassy meadow surrounded by forest hills, wove the adventurous and playful snake named Oscar. He was brown and yellow in colour, with a white diamond pattern on his back.

Oscar was always looking for new places to explore, so one day he decided to wander through the forest hills. He slithered up and down the dirt paths, taking in all of nature's beauty around him.
As he kept moving, Oscar noticed a small pond in the middle of the forest. He decided to take a closer look and when he got there, he was in awe. The pond was crystal clear and filled with lily pads and colorful fish.

Oscar couldn't resist the temptation and he quickly jumped into the water. He swam around for hours, enjoying the cool and refreshing sensation of the pond's waters.

As the sunlight dipped behind the trees, Oscar decided to pay one last exploration visit before he finished up for the day. As he rounded a corner, he came face to face with a giant boa constrictor. The two were both startled, but the boa quickly wrapped itself around Oscar and squeezed tight.

Oscar was scared and he knew that his life was in danger, but then something amazing happened. The boa constrictor loosened its strong grip and released Oscar without hurting him. The two then just looked at each other, as if they were communicating something neither of them could understand.

Oscar was so surprised by the snake's kindness and he knew that this experience had changed him forever. From then on,t the world around him was full of wonders, just waiting for him to discover. 

\chapter{Log Data}
\section{Sample Log Data 1 (Game Version P11))}
[3/6/2023 7:29:47 PM]Game Start

[3/6/2023 7:29:48 PM][1][3]Once lived in an old abandoned house in the center of a busy city, was a black-bellied golden snake. Day by day he would
[3/6/2023 7:29:48 PM][0][1]Once upon a time there was a snake named Slither who lived in the forest. He had been living peacefully and happily until one day, when

[3/6/2023 7:30:03 PM]Chose[1][3]

[3/6/2023 7:30:04 PM][1][0] scour up and down the crevices of this broken home unnoticed. Occasionally it crawled out into the scared city ridden society to cultivate a small hoard
[3/6/2023 7:30:04 PM][0][1] slither around the house, exploring every corner and crevice.

One day he was out for a stroll in the garden of the old house
[3/6/2023 7:30:17 PM][2][5]

[3/6/2023 7:30:21 PM]Chose[1][0]

[3/6/2023 7:30:22 PM][1][3] of food just enough to keep warm. As chilling winter winds slowly rolled in and darkness increased, no bouts of sunlight hit the grimy abode itself
[3/6/2023 7:30:23 PM][0][1] of food. 

One day, the snake decided to venture out a bit further into the city than usual. He was mesmerized by all

[3/6/2023 7:30:52 PM]Chose[1][3]

[3/6/2023 7:30:53 PM][1][4]. 

Seasons unearthed happiness all around this cozy old home, kids screeched and teenage gossip shook the foundation undoubtedly our friend welcomed shelter in
[3/6/2023 7:30:53 PM][0][1].

The snake was quite content with this lifestyle until one day, when it heard a strange noise coming from the house. Curiosity got the better

[3/6/2023 7:31:19 PM]Chose[0][1]

[3/6/2023 7:31:21 PM][1][3] of it and so, he slithered its way inside towards the commotion. To his surprise were two other snakes - one bright orange spotted snake that
[3/6/2023 7:31:21 PM][0][2] of it, and the snake decided to investigate. As it slithered its way up the stairs, towards the source of this noise, its scales began

[3/6/2023 7:32:09 PM]Chose[0][2]

[3/6/2023 7:32:10 PM][1][4] a tapestry of gold and red waves in the sometimes pitch twilight. When finally it made its way to place from where this racket fell on its
[3/6/2023 7:32:10 PM][0][0] to tremble in fear. It was a human, and it seemed as if he had found the snake's secret hideout.

The man

[3/6/2023 7:32:44 PM]Chose[1][4]

[3/6/2023 7:32:46 PM][0][2] ears, it was astounded by what it saw.

There in the center of this dingy room stood a small boy, no older than
[3/6/2023 7:32:46 PM][1][4] ears, what it encountered swore changed his life forever.

Two people lay huddled next to in the interlocked embrace; poor souls everywhere had
[3/6/2023 7:33:12 PM][2][5]he snake had a massive concussion and it couldn't move for a few minutes.

[3/6/2023 7:33:16 PM]Chose[2][5]

[3/6/2023 7:33:17 PM][0][0] 

When it finally regained its senses, the snake realized that the noise was coming from a group of people who were renovating this old home
[3/6/2023 7:33:18 PM][1][0] To it surprise the intruder was an orange rat, thief like eyes darting to and fro then, looking at it . On its left hip pocket was

[3/6/2023 7:34:39 PM]Chose[0][0]

[3/6/2023 7:34:40 PM][1][4]. It had instantly scored itself an arduous task - Surviving this hostile environment would take hidden amidst the remodelling mess- all depending on patience and
[3/6/2023 7:34:40 PM][0][2]. They had come to give the house a much needed facelift and were unearthing all its hidden secrets. However, in their haste they

[3/6/2023 7:35:18 PM]Chose[1][4]

[3/6/2023 7:35:19 PM][0][2] luck. 

The snake, however, was a master of unseen and it managed to stay hidden for months on end until the house was finally
[3/6/2023 7:35:19 PM][1][0] strength alike.
Evey night while evading punches and hammer-killer blows, that very same unsettling snake somehow managed to maintain existence into this confusing
[3/6/2023 7:35:44 PM][2][5]and it's agility of movement.

[3/6/2023 7:35:49 PM]Chose[2][5]

[3/6/2023 7:35:50 PM][0][0]The snake was able to dodge the workers' tools and their careless feet. For weeks, it stayed hidden in a corner of the house,
[3/6/2023 7:35:51 PM][1][4] Determined to beat the odds, and escape both unscathed plus undestroyed the snake definitely nested deep down regular ceilings and walls. Now is

[3/6/2023 7:36:54 PM]Chose[1][4]

[3/6/2023 7:36:56 PM][0][1] the time to wait and watch. 

As days passed, the snake became accustomed to its new home - it knew where it could find food
[3/6/2023 7:36:56 PM][1][0] another member of new house happily nestled furniture plots manoeuvring escape for various protected corners. Growing about doubled its apparel and authority, this dust-rake
[3/6/2023 7:37:24 PM][2][5]ind it's way down a pipe and into a safe underground haven. Aka the sewer.

[3/6/2023 7:37:27 PM]Chose[2][5]

[3/6/2023 7:37:28 PM][0][2] 

The snake stayed there for many years, while the house was slowly remodelled and restored to its former glory. As time passed by,
[3/6/2023 7:37:28 PM][1][0]Gearing up each weakening suna with newfoundfault rather vighiand swimming tamanringly tyrying es never laidoutsdong

[3/6/2023 7:37:45 PM]Chose[1][0]

[3/6/2023 7:37:47 PM][0][2] in the sewers, learning to live off of whatever edible morsels it could find. 

Years passed and the snake slowly adapted to this
[3/6/2023 7:37:47 PM][1][0]obtained ulumizedeach tiny wramping word ways falwen curprise lifhpiful second uphew.Finally iazuplettsand w

[3/6/2023 7:38:13 PM]Chose[1][0]

[3/6/2023 7:38:15 PM][1][3]all waters an passiffirece teleasight immish are found campfty confidescedi hraculumsetll tearchedting old for
[3/6/2023 7:38:15 PM][0][1]hile the snake could barely remember it's old life, it was content to spend its days in the sewers of the city.

[3/6/2023 7:38:29 PM]Game End

[3/6/2023 7:38:29 PM]Ate[14]

\section{Sample Log Data 1 (Non-Game Version P11))}

[3/6/2023 7:49:53 PM]Game Start

[3/6/2023 7:49:54 PM][0.6]Once upon a time there was a small snake who lived in the forest. The snake loved to explore his surroundings and would often slither around looking
[3/6/2023 7:49:54 PM][1.4]When Gus came to visit his uncle's country farm he was excited for the adventures that awaited him. He had heard all about how this new environment offered

[3/6/2023 7:50:22 PM][Add Own Text]There once was a Boy named George that lived on a farm near a forest.

[3/6/2023 7:50:23 PM][0.6] One day, while walking in the forest, he came across a large snake. The snake was coiled up and had bright yellow eyes that seemed to
[3/6/2023 7:50:23 PM][1.4] On the very edge of what felt like a forbidden wild abyss, there was an old abandoned shed that felt particularly displaced from Nature. Little did George know

[3/6/2023 7:50:54 PM][Chose][0.6]

[3/6/2023 7:50:55 PM][0.6] look right through him.

George was scared, but he knew that snakes were important to the environment and so he decided not to run away.
[3/6/2023 7:50:55 PM][1.4] stare straight into George's soul. As George stepped closer, his eyes began to widen as fear and tension filled the air.

George hesitated at

[3/6/2023 7:51:34 PM][Add Own Text]e was now convinced that he was chosen to be a member of the dark side.

[3/6/2023 7:51:35 PM][0.6] He knew that he could no longer live his old life and had to accept the snake as his master. 

He followed the snake deep into
[3/6/2023 7:51:35 PM][1.4]He timidly reached out to softly touch its flat yet slimy black color. When he had finally worked up enough courage, his finger immediately

[3/6/2023 7:51:49 PM][Chose][0.6]

[3/6/2023 7:51:50 PM][0.6] the forest and to a large clearing with an old, decrepit cabin. The snake slithered around George and motioned for him to enter the cabin
[3/6/2023 7:51:50 PM][1.4] the Forest and before long, they came across an old temple. Inside the temple was a large room filled with pits of glowing green goo - this

[3/6/2023 7:52:31 PM][Chose][1.4]

[3/6/2023 7:52:32 PM][0.6] was the snake's home. 

The snake spoke to George, telling him that he could have anything he wanted if he pledged his allegiance and
[3/6/2023 7:52:33 PM][1.4] was no ordinary sight. The snake slithered in and placed itself at the centre of the room before beginning to speak:

"Close humanling

[3/6/2023 7:53:18 PM][Chose][1.4]

[3/6/2023 7:53:19 PM][0.6], I am the guardian of this temple and you have been chosen to serve me. From now on you will be my servant and do as I command
[3/6/2023 7:53:20 PM][1.4], it's time ye became one with the snakes of dragons. I shall be tay tupping hearfy to protect youalways from lives touched barren

[3/6/2023 7:53:39 PM][Chose][1.4]

[3/6/2023 7:53:40 PM][0.6]. I grant ye the power to control the elements of this world, but only when you accept my blessing."

George was terrified and uncertain what
[3/6/2023 7:53:40 PM][1.4]. Gallivanting about every daye frivolipaseday joorneys with on larkining flightal skeemeshyne lizard wingable sprites

[3/6/2023 7:54:05 PM][Chose][1.4]

[3/6/2023 7:54:06 PM][0.6]. 

George was amazed by what he heard and felt compelled to accept the snake's offer, so he nodded in agreement. He quickly began
[3/6/2023 7:54:06 PM][1.4]. Forewards danwortion glymback bowards blessid them plateary putenderbound beasts."

George was astounded but knew this

[3/6/2023 7:54:12 PM][Chose][1.4]

[3/6/2023 7:54:14 PM][0.6] was his destiny. He bowed down before the snake and accepted its offer to become a part of its family. From then on, George became known as
[3/6/2023 7:54:14 PM][1.4] must be the purpose of whatever journey lay ahead he was been. The snake placed its hood across his head and instructed him ingest a mysterious blue liquid in

[3/6/2023 7:54:44 PM][Add Own Text]absolute gibberish must be a forgotten tongue spoken by the ancient snake

[3/6/2023 7:54:45 PM][1.4]s of dragons. With a slight belly flutter, he stepped out with the Giant Snake behind him and declared his loyalty to protecting its power full force
[3/6/2023 7:54:45 PM][0.6]s. He nodded his head in agreement and from that day forward, became the snake's apprentice. The snake taught George the forgotten language of dragons and

[3/6/2023 7:55:17 PM][Chose][0.6]

[3/6/2023 7:55:19 PM][1.4] came to see him as a son. George worked hard learning the magicks from his master and embracing the darkness with open arms and righteous judgment.

[3/6/2023 7:55:19 PM][0.6] together they travelled to distant lands, learning secrets and exploring the unknown. 

Years passed by and George became an expert in dragon lore and a

[3/6/2023 7:55:46 PM][Chose][1.4]

[3/6/2023 7:55:47 PM][0.6]Years later, George had become a powerful wizard and a respected leader of the dark side. He was feared by many but respected for his wisdom and
[3/6/2023 7:55:48 PM][1.4]When George reached maturity, he joined the warrior army of Serpentin Graywitch who worshipped him and regarded him as one of their champions. George made

[3/6/2023 7:56:06 PM][Chose][1.4]

[3/6/2023 7:56:07 PM][1.4] excellent achievements in the army, never losing a battle or mission. Little by little, with the help of George its prime warrior, the dark forces slowly
[3/6/2023 7:56:07 PM][0.6] sure to always keep the snake close by his side and together they fought off many enemies.

The snake was very proud of George and eventually rewarded

[3/6/2023 7:56:25 PM][Chose][1.4]

[3/6/2023 7:56:26 PM][0.6] began to take over the world. 

Today, George is known as the King of Snakes and his faithful servant still by his side.
[3/6/2023 7:56:27 PM][1.4] took control of the entire landscape and drew all things into itselfy. Who knew one random meeting with a snake in the forest would lead to such power

[3/6/2023 7:56:44 PM][Chose][1.4]

[3/6/2023 7:56:45 PM][0.6]!
[3/6/2023 7:56:45 PM][1.4]?

[3/6/2023 7:56:47 PM][Chose][1.4]

[3/6/2023 7:57:01 PM]Game End

\chapter{Survey Questions}

\section{Story Evaluation}
In a degree of 5 (from strongly disagree to strongly agree), answer the following questions:
\begin{itemize}
    \item [Q1] I think the logic of the story is well organized.
    \item [Q2] I think the language in the story was professionally written.
    \item [Q3] I think the overall quality of the story was perfect.
\end{itemize}

\section{Experience Evaluation}
In a degree of 5 (from strongly disagree to strongly agree), answer the following questions:
\begin{itemize}
    \item [Q1] I think the story is written by myself.
    \item [Q2] I think I prioritize the quality of the story in the game.
    \item [Q3] I think I like this story and want to share it with others.
    \item [Q1] I think the system is too complex to be understood.
    \item [Q2] I think the gameplay interrupts my thinking while writing the story.
    \item [Q3] I think I am engaged in the co-writing process.
\end{itemize}

\section{Demographic Questions}
\begin{itemize}
    \item [Q1] Did you co-create content with any Artificial Intelligence before? Y/N
    \item [Q2] How would you describe your writing skills? 
    \begin{itemize}
        \item [1] Never wrote stories
        \item [2] Have some skeletons of stories but never complete them
        \item [3] Wrote some stories and shared them with others privately or published them
    \end{itemize}
 
\end{itemize}

\printindex

\end{document}